\documentclass[pra,aps,showpacs,longbibliography,twocolumn]{revtex4}
\UseRawInputEncoding

\usepackage{amssymb}
\usepackage{amsmath}
\usepackage{mathtools}
\usepackage[dvipdfm]{hyperref}
\usepackage{hyperref}

\usepackage{bm}
\usepackage{graphicx}
\usepackage{mathrsfs}

\allowdisplaybreaks
\begin{document}

\title{Estimating coherence with respect to general quantum measurements}
\author{Jianwei Xu\textsuperscript{1}}
\email{xxujianwei@nwafu.edu.cn}
\author{Lin Zhang\textsuperscript{2}}
\email{linyz@hdu.edu.cn}
\author{Shao-Ming Fei\textsuperscript{3,4}}
\email{feishm@cnu.edu.cn}
\affiliation{\textsuperscript{1}College of Science, Northwest A\&F University, Yangling, Shaanxi 712100,
China}
\affiliation{\textsuperscript{2}Institute of Mathematics, Hangzhou Dianzi University, Hangzhou 310018, China}
\affiliation{\textsuperscript{3}School of Mathematical Sciences, Capital Normal University, Beijing 100048, China}
\affiliation{\textsuperscript{4}Max-Planck-Institute for Mathematics in the Sciences, Leipzig 04103, Germany}

%\date{\today }

\begin{abstract}

The conventional coherence is defined with respect to a fixed
orthonormal basis, i.e., to a von Neumann measurement.
Recently, generalized quantum coherence with respect to general positive operator-valued measurements
(POVMs) has been presented. Several well-defined coherence measures, such as the relative entropy
of coherence $C_{r}$, the $l_{1}$ norm of coherence $C_{l_{1}}$ and the
coherence $C_{T,\alpha }$ based on Tsallis relative entropy with respect to general POVMs
have been obtained. In this work, we investigate the properties of $C_{r}$, $l_{1}$ and $C_{T,\alpha }$. We estimate the upper bounds of $C_{l_{1}}$; we show that
the minimal error probability of the least square measurement state discrimination is given by $C_{T,1/2}$; we derive the uncertainty relations given by $C_{r}$, and calculate the
average values of $C_{r}$, $C_{T,\alpha }$ and $C_{l_{1}}$ over random pure
quantum states. All these results include the corresponding results of the conventional
coherence as special cases.

\end{abstract}

\pacs{03.65.Ud, 03.67.Mn, 03.65.Aa}
\maketitle

\section{Introduction}

Coherence is a fundamental feature in quantum physics and a significant resource in
quantum information processing. Since the rigorous framework \cite{BCP-2014-PRL} for quantifying coherence has been established, fruitful advances about coherence, both in theories and experiments, have been achieved, for reviews see e.g. \cite{Plenio-2016-RMP,Fan-2018-PhysicsReports}.
The conventional coherence is defined with respect to a fixed orthonormal
basis. We know that an orthonormal basis $\{|j\rangle \}_{j=1}^{d}$ of a $d$-dimensional complex Hilbert space $H$,
corresponds to a rank-one projective measurement (von Neumann measurement) with measurement operators $\{|j\rangle \langle j|\}_{j=1}^{d}$. Along this line, we may ask whether or not the notion of coherence can be generalized to general measurements. Recently, Bischof, Kampermann and Bru${\ss}$ generalized the conventional framework of coherence
to the case of general positive operator-valued measurements (POVMs), by replacing the projective measurements
with POVMs \cite{Brub-2019-PRL,Brub-2021-PRA}.

Let $\mathcal{S}(H)$ be the set of all quantum states on $H,$ $%
\mathcal{C}(H)$ be the set of all channels on $H$ \cite{Nielsen-2000-book}.
For a POVM $E=\{E_{j}\}_{j=1}^{n}$, $E_{j}\geq 0$ $\forall j$ and $\sum_{j=1}^{n}E_{j}=I_{d} $ with $I_{d}$ the identity on $H$,
the set of POVM incoherent states is specified as \cite{Brub-2019-PRL}
\begin{eqnarray}
\mathcal{I}_{\text{p}}(H)=\{\rho \in \mathcal{S}(H)|E_{j}\rho E_{k}=0,~ \forall  j\neq k \}.    \label{eq1.1}
\end{eqnarray}
A channel $\phi \in \mathcal{C}(H)$ is called a POVM incoherent channel \cite{Brub-2021-PRA} if
$\phi $ has a Kraus operator decomposition $\phi =\{K_{l}\}_{l}$ with $\sum_{l}K_{l}^{\dag }K_{l}=I_{d}$
and there exists a block incoherent channel $\{K_{l}^{\prime }\}_{l}$ with
respect to a canonical Naimark extension such that
\begin{eqnarray}
K_{l}\rho K_{l}^{\dag }\otimes |1\rangle \langle 1|=K_{l}^{\prime }(\rho
\otimes |1\rangle \langle 1|)K_{l}^{^{\prime }\dag },~\forall \ l,   \label{eq1.2}
\end{eqnarray}
where $\dag $ stands for conjugate transpose. We call such a decomposition $%
\phi =\{K_{l}\}_{l}$ a POVM incoherent decomposition. We denote by $\mathcal{C}_{%
\text{PI}}(H)$ the set of all POVM incoherent channels.

A functional $C:\mathcal{S}(H)\rightarrow R$ is called a POVM coherence
measure with respect to the POVM $E=\{E_{j}\}_{j=1}^{n}$ if $C$ fulfills the
following conditions \cite{Brub-2021-PRA}.

(P1) Faithfulness: $C(\rho,E)\geq 0$, with equality if and only if $\rho\in
\mathcal{I}_{\text{P}}(H)$.

(P2) Monotonicity: $C(\phi _{\text{PI}}(\rho),E)\leq C(\rho,E),$ $%
\forall $ $\phi _{\text{PI}}\in \mathcal{C}_{\text{PI}}(H)$.

(P3) Strong monotonicity: $\sum_{l}p_{l}C(\rho _{l},E)\leq C(\rho,E),$
$\forall $ $\phi _{\text{PI}}\in \mathcal{C}_{\text{PI}}(H),$
where $\phi _{\text{PI}}=\{K_{l}\}_{l}$ is a POVM incoherent decomposition
of $\phi _{\text{PI}}$, $p_{l}=$tr$(K_{l}\rho K_{l}^{\dag }),\rho
_{l}=K_{l}\rho K_{l}^{\dag }/p_{l}$.

(P4) Convexity: $C(\sum_{j}p_{j}\rho _{j},E)\leq \sum_{j}p_{j}C(\rho
_{j},E),$ $\{\rho _{j}\}_{j}\subset \mathcal{S}(H),$ $\{p_{j}\}_{j}$ a probability
distribution.

When the POVM $E$ is a rank-one projective measurement $%
E=\{|j\rangle \langle j|\}_{j=1}^{d}$, the definitions of POVM incoherent
states and POVM incoherent channels, as well as the conditions (P1) to (P4) all
reduce to the cases of the conventional coherence theory, for which various kinds of coherence measures have
been proposed together with their operational interpretations and applications
\cite{BCP-2014-PRL,XY,AW,CN,BC,KB1,Winter-2017-PRSA,Yu-2017-PRA,Yu-2018-SR,Xiong-2018-PRA,LUO8,LUO9,LUO10,Xu-2020-CPB,XNZ}.
However, less is known about the POVM coherence measures. Recently,
several POVM coherence measures have been proposed, such as robustness of POVM coherence
$C_{\text{rob}}(\rho ,E)$ \cite{Brub-2021-PRA}, $%
l_{1}$-norm of POVM coherence $C_{l_{1}}(\rho ,E)$ \cite{Brub-2021-PRA,Xu-2020-PRA}, relative entropy of POVM coherence $%
C_{r}(\rho ,E)$ \cite{Brub-2019-PRL,Brub-2021-PRA}, and POVM coherence $C_{T,\alpha }(\rho ,E)$ based on Tsallis
relative entropy \cite{Xu-2020-PRA}. In particular, the $C_{r}(\rho ,E)$, $C_{l_{1}}(\rho ,E)$ and $C_{T,\alpha }(\rho ,E)$ allow for explicit  expressions. Furthermore, an alternative definition of POVM incoherent operation was introduced \cite{Wu-2021-PRA}, and the quantifiers of POVM coherence based on max-relative entropy and coherent rank were studied \cite{Gao-2021-arxiv}.

$C_{r}(\rho ,E)$ is defined by \cite{Brub-2019-PRL,Brub-2021-PRA},
\begin{eqnarray}
C_{r}(\rho ,E)=\sum_{j}S(\sqrt{E_{j}}\rho \sqrt{E_{j}})-S(\rho ),  \label{eq1.3}
\end{eqnarray}
where $S(M)=-$tr$(M\log _{2}M)$ is the entropy for positive semidefinite matrix $M.$

$l_{1}$-norm of POVM coherence $C_{l_{1}}(\rho ,E)$ \cite{Brub-2021-PRA,Xu-2020-PRA} is defined as
\begin{eqnarray}
C_{l_{1}}(\rho ,E)=\sum_{j\neq k}||\sqrt{E_{j}}\rho \sqrt{E_{k}}||_{\text{tr%
}},  \label{eq1.4}
\end{eqnarray}
with $||M||_{\text{tr}}=$tr$\sqrt{M^{\dag }M}$ the trace norm of matrix $M.$

 $C_{T,\alpha }(\rho,E)$ is defined by \cite{Xu-2020-PRA}
\begin{eqnarray}
C_{T,\alpha }(\rho ,E)=\frac{1}{\alpha -1}\{\sum_{j}\text{tr}[(\sqrt{E_{j}}\rho
^{\alpha }\sqrt{E_{j}})^{1/\alpha }]-1\} \label{eq1.5}
\end{eqnarray}
for $\alpha \in (0,1)\cup (1,2]$.
When $E$ is a rank-one projective measurement, $C_{T,\alpha }(\rho,E)$ returns to the standard coherence measure $C_{r}(\rho ,\{|j\rangle \langle j|\}_{j=1}^{d})$ proposed in Ref. \cite{Xiong-2018-PRA,Yu-2018-SR}. A quantifier of standard coherence measure based on Tsallis entropy was proposed in Ref. \cite{Rastegin-2016-PRA}, however it does not satisfy all conditions of BCP framework.

In this work, we investigate the properties and the estimations of $C_{r}(\rho ,E)$, $C_{l_{1}}(\rho ,E)$ and $C_{T,\alpha }(\rho ,E)$.
We estimate the upper bounds and the averages for random pure quantum states, and explore their operational interpretations
and the related uncertainty relations, which strengthen the
necessity and applicability of the coherence with respect to POVMs. This paper is organized as
follows. In section II, we discuss the upper bounds for $C_{l_{1}}(\rho ,E).$
In section III, we provide an operational interpretation for $C_{T,1/2}(\rho
,E).$ In section IV, we establish an uncertainty relation for $C_{r}(\rho
,E). $ In section V, we calculate the average values of $C_{l_{1}}(\rho
,E)$, $C_{r}(\rho ,E)$ and $C_{T,\alpha }(\rho ,E)$ for random pure states.
Section VI is a brief summary.

\section{Upper bounds for $C_{l_{1}}$}
We first estimate the upper bounds of $C_{l_{1}}(\rho ,E)$.
From the Theorem 4.29 in \cite{Zhan-book-2002}, we have that for any positive semidefinite
matrices $A,B$ and matrix $X$,
\begin{eqnarray}
||AXB||_{\text{tr}}\leq ||A^{p}X||_{\text{tr}}^{1/p}||XB^{q}||_{\text{tr}%
}^{1/q} \label{eq2.2}
\end{eqnarray}
for all positive numbers $p>1$, $q>1$ and $1/p+1/q=1.$
From (\ref{eq2.2}) we obtain
\begin{eqnarray}
||\sqrt{E_{j}}\rho \sqrt{E_{k}}||_{\text{tr}}\leq ||E_{j}^{p/2}\rho ||_{%
\text{tr}}^{1/p}||E_{k}^{q/2}\rho ||_{\text{tr}}^{1/q}. \label{eq2.3}
\end{eqnarray}
Summing over all $j\neq k$ we get the following Theorem.

\emph{Theorem 1.} For any POVM $E=\{E_{j}\}_{j=1}^{n}$ and quantum state $\rho$ it holds that
\begin{eqnarray}
C_{l_{1}}(\rho ,E)\leq \sum_{j\neq k}||E_{j}^{p/2}\rho ||_{\text{tr}%
}^{1/p}||E_{k}^{q/2}\rho ||_{\text{tr}}^{1/q},  \label{eq2.1}
\end{eqnarray}
where $p>1$, $q>1$ and $1/p+1/q=1$.

For $j\neq k$, one sees that from (\ref{eq1.4}), $C_{l_{1}}(\rho,E)$
is dependent on both the state $\rho$ and the commutativity of $\{E_{j},E_{k}\}$
and $\rho$. However, the upper bound $||E_{j}^{p/2}\rho ||_{\text{tr}%
}^{1/p}||E_{k}^{q/2}\rho ||_{\text{tr}}^{1/q}$ is a factorized form of $%
E_{j} $ over $\rho $ and $E_{k}$ over $\rho$. Theorem 1 sets a constraint
on such factorization.

When $p=q=2$, (\ref{eq2.1}) becomes
\begin{eqnarray}
C_{l_{1}}(\rho ,E)&\leq& \sum_{j\neq k}||E_{j}\rho ||_{\text{tr}%
}^{1/2}||E_{k}\rho ||_{\text{tr}}^{1/2}  \nonumber \\
&=&(\sum_{j}||E_{j}\rho ||_{\text{tr}}^{1/2})^{2}-\sum_{j}||E_{j}\rho ||_{%
\text{tr}}.  \label{eq2.4}
\end{eqnarray}
When $p\rightarrow 1$ and $q\rightarrow +\infty$, we have Theorem 2 below.

\emph{Theorem 2.} For any POVM $E=\{E_{j}\}_{j=1}^{n}$ and quantum state $\rho$, we have
\begin{eqnarray}
&&C_{l_{1}}(\rho ,E) \nonumber \\
&\leq& 2\sum_{j}(n-j)||(E_{j}^{1/2}\rho )^{\uparrow }||_{\text{tr}}   \label{eq2.6} \\
&\leq& (n-1)\sum_{j}||E_{j}^{1/2}\rho ||_{\text{tr}}, \label{eq2.7}
\end{eqnarray}
where $\{(E_{j}^{1/2}\rho )^{\uparrow }\}$ stands for $(E_{1}^{1/2}\rho )\leq
(E_{2}^{1/2}\rho )\leq ...\leq (E_{n}^{1/2}\rho )$.

\emph{Proof.} Let
$E_{k}=\sum_{l_{k}}\lambda _{k,l_{k}}^{\downarrow }I(\lambda _{k,l_{k}})$
be the eigendecomposition of $E_{k},$ where $I(\lambda _{k,l_{k}})$ is the
identity operator of the subspace spanned by all eigenvectors corresponding to the
eigenvalues $\lambda_{l_{k}}$, $\{\lambda _{k,l_{k}}^{\downarrow }\}_{l_{k}}$ stands for
$\lambda_{k,1}>\lambda _{k,2}>...$, arranged in strictly decreasing order.
Consequently, we get
\begin{eqnarray}
E_{k}^{q/2}&=&\sum_{l_{k}}(\lambda _{k,l_{k}}^{\downarrow })^{q/2}I(\lambda
_{k,l_{k}}),  \nonumber \\
||E_{k}^{q/2}\rho ||_{\text{tr}}&=&\text{tr}\sqrt{\rho E_{k}^{q}\rho } \nonumber \\
&\leq& \text{tr}
\sqrt{\rho \lbrack \chi (E_{k})]^{q}I_{d}\rho }=[\chi (E_{k})]^{q/2}, \nonumber \\
||E_{k}^{q/2}\rho ||_{\text{tr}}^{1/q}&\leq& \sqrt{\chi (E_{k})},\nonumber
\end{eqnarray}
where
\begin{eqnarray}
\chi (E_{k})=
\begin{cases}
\max \{\lambda \in \sigma (E_{k})|I(\lambda )\rho
\neq 0\},~\text{when}~ \ E_{k}\neq 0, \\
0, ~\text{when}~ \ E_{k}=0,
\end{cases}\nonumber
%\label{eq2.8}
\end{eqnarray}
$\sigma (E_{k})$ is the set of eigenvalues of $E_{k}$.

When $p\rightarrow 1$ and $q\rightarrow +\infty$, we have $||E_{j}^{p/2}\rho ||_{%
\text{tr}}^{1/p}\rightarrow ||E_{j}^{1/2}\rho ||_{\text{tr}}$. Therefore, from Theorem 1,
$C_{l_{1}}(\rho ,E)\leq \sum_{j\neq k}\min \{||E_{j}^{1/2}\rho ||_{\text{tr}}
\sqrt{\chi (E_{k})},||E_{k}^{1/2}\rho ||_{\text{tr}}\sqrt{\chi (E_{j})}\}$.
Since $E_{k}^{q/2}\leq I_{d},$ we get $\sqrt{\chi (E_{k})}\leq 1$, and
\begin{eqnarray}
C_{l_{1}}(\rho ,E)&\leq& \sum_{j\neq k}\min \{||E_{j}^{1/2}\rho ||_{\text{tr}%
},||E_{k}^{1/2}\rho ||_{\text{tr}}\}   \nonumber \\
&=&\sum_{j\neq k}\min \{||(E_{j}^{1/2}\rho )^{\uparrow }||_{\text{tr}%
},||(E_{k}^{1/2}\rho )^{\uparrow }||_{\text{tr}}\}   \nonumber \\
&=&2\sum_{j<k}\min \{||(E_{j}^{1/2}\rho )^{\uparrow }||_{\text{tr}%
},||(E_{k}^{1/2}\rho )^{\uparrow }||_{\text{tr}}\}  \nonumber \\
&=&2\sum_{j=1}^{n}(\sum_{k=j+1}^{n}||(E_{j}^{1/2}\rho )^{\uparrow }||_{\text{tr}})   \nonumber \\
&=&2\sum_{j}(n-j)||(E_{j}^{1/2}\rho )^{\uparrow }||_{\text{tr}},
\end{eqnarray}
which gives the inequality (\ref{eq2.6}).

From the definition of $\{||(E_{j}^{1/2}\rho )^{\uparrow }||_{\text{tr}}\},$ we have
\begin{eqnarray}
(n-j)||(E_{j}^{1/2}\rho )^{\uparrow }||_{\text{tr}}\leq
\sum_{l=j+1}^{n}||(E_{l}^{1/2}\rho )^{\uparrow }||_{\text{tr}}.   \label{eqA1.2}
\end{eqnarray}
Adding $(n-j)||(E_{j}^{1/2}\rho )^{\uparrow }||_{\text{tr}}$ to both sides of
(\ref{eqA1.2}) and summing over the index $j$, we obtain
\begin{eqnarray}
2\sum_{j=1}^{n}(n-j)||(E_{j}^{1/2}\rho )^{\uparrow }||_{\text{tr}}   \ \ \ \ \ \ \ \  \ \ \ \ \ \ \ \     \ \ \ \ \ \ \ \  \ \ \ \ \ \ \ \   \ \ \ \    \nonumber \\
\leq
\sum_{j=1}^{n}\sum_{l=j+1}^{n}||(E_{l}^{1/2}\rho )^{\uparrow }||_{\text{tr}%
}
+\sum_{j=1}^{n}(n-j)||(E_{j}^{1/2}\rho )^{\uparrow }||_{\text{tr}}. \ \ \ \   \label{eqA1.3}
\end{eqnarray}
Since the first term in (\ref{eqA1.3}),
\begin{eqnarray}
\sum_{j=1}^{n}\sum_{l=j+1}^{n}||(E_{l}^{1/2}\rho )^{\uparrow }||_{\text{tr}%
}=\sum_{j=1}^{n}(j-1)||(E_{j}^{1/2}\rho )^{\uparrow }||_{\text{tr}},\nonumber
\end{eqnarray}
(\ref{eqA1.3}) leads to (\ref{eq2.7}). $\hfill\blacksquare$

When $E$ is a rank-one projective measurement $%
E=\{|j\rangle \langle j|\}_{j=1}^{d}$, we have the following Corollary from the inequality (\ref{eq2.6}) in Theorem 2.

\emph{Corollary 1.} For rank-one projective measurement $E=\{|j\rangle \langle
j|\}_{j=1}^{d},$ and any quantum state $\rho ,$ we have
\begin{eqnarray}
C_{l_{1}}(\rho ,\{|j\rangle \langle j|\}_{j=1}^{d})\leq 2\sum_{j}(d-j)\sqrt{%
\langle j|\rho ^{2}|j\rangle }^{\uparrow }\equiv B_{1},  \label{eq2.9}
\end{eqnarray}
where $\{\sqrt{\langle j|\rho ^{2}|j\rangle }^{\uparrow }\}_{j}$ stands for arranging
$\{\sqrt{\langle j|\rho ^{2}|j\rangle }\}_{j}$ in nondecreasing order.

In Ref. \cite{Chen-PRA-2015} the authors presented
upper bounds for $C_{l_{1}}(\rho ,\{|j\rangle \langle j|\}_{j=1}^{d})$, see Eqs. (2,7) in Ref. \cite{Chen-PRA-2015},
\begin{eqnarray}
&&C_{l_{1}}(\rho ,\{|j\rangle \langle j|\}_{j=1}^{d})\leq (\sum_{j}\sqrt{%
\langle j|\rho |j\rangle })^{2}-1\equiv B_{2}, \ \ \ \ \ \  \label{eq2.10}  \\
&&C_{l_{1}}(\rho ,\{|j\rangle \langle j|\}_{j=1}^{d})   \nonumber \\
&\leq& \sqrt{d(d-1)(\text{%
tr}\rho ^{2}-\sum_{j}\langle j|\rho |j\rangle ^{2})}\equiv B_{3}.  \label{eq2.11}
\end{eqnarray}
We compare the upper bound $B_{1}$ with the upper bounds $B_{2}$ and
$B_{3}$ by detailed examples.

\emph{Example 1.} Consider state
\begin{eqnarray}
\rho =\frac{1}{2}\left(
\begin{array}{cc}
1-z & \frac{1}{2} \\
\frac{1}{2} & 1+z%
\end{array}%
\right),~~~z\in \lbrack 0,\frac{4}{5}],  \label{eq2.12}
\end{eqnarray}
under $d=2$ orthonormal basis $\{|j\rangle\}_{j=1}^{2}$. We have
\begin{eqnarray}
B_{1}&=&\sqrt{\frac{1}{4}+(1-z)^{2}},  \nonumber \\
B_{2}&=&\sqrt{1-z^{2}}, \nonumber \\
B_{3}&=&\frac{1}{2}.  \nonumber
\end{eqnarray}
We see that $B_{1}$ is tighter than $B_{2}$ for certain states.
In particular, when $z=0.5$ we have $B_{2}>B_{1}>B_{3};$ while when $z=0.1,$ $B_{1}>B_{2}>B_{3}$, see Fig. 1.

\emph{Example 2.} Consider $\rho =|\psi \rangle \langle \psi |$, with
\begin{eqnarray}
|\psi \rangle =x|1\rangle +4x|2\rangle +\sqrt{1-17x^{2}}|3\rangle, \label{eq2.13}
\end{eqnarray}
where $x\in\lbrack 0,\frac{1}{\sqrt{17}}]$, $\{|j\rangle\}_{j=1}^{3}$ is an orthonormal basis. One has
\begin{eqnarray}
B_{1}&=&12x,  \nonumber \\
B_{2}&=&(5x+\sqrt{1-17x^{2}})^{2}-1, \nonumber \\
B_{3}&=&\sqrt{6[1-17x^{4}-(1-17x^{2})^{2}]}.
\end{eqnarray}
One sees that when $x=0.1,$ $%
B_{2}<B_{1}<B_{3};$ while when $x=0.21,$ $B_{2}<B_{3}<B_{1}$, see Fig. 2.

\begin{figure}[!h]
\includegraphics[width=8cm]{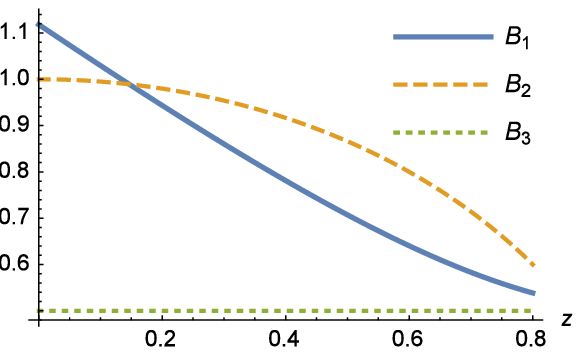}
\caption{$B_{1}$, $B_{2}$ and $B_{3}$ versus $z$ for the states in Example 1.}
~\\
\includegraphics[width=8cm]{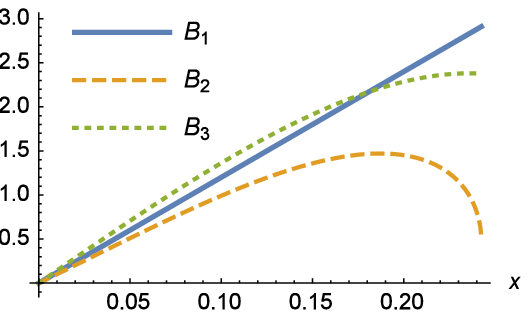}
\caption{$B_{1}$, $B_{2}$ and $B_{3}$ versus $x$ for the states in Example 2.}
\end{figure}

Example 1 and Example 2 show that the upper bound $B_{1}$ is a new bound different from $B_{2}$ and $B_{3}$, and $B_{1},B_{2},B_{3}$ have no strict order of which is tighter than others.

\section{Operational interpretation of $C_{T,1/2}$}
In this section, we provide an operational interpretation for $%
C_{T,1/2}.$ We first review the least square measurement (LSM). Quantum
state discrimination is a fundamental problem in quantum physics, and
plays a key role in quantum communication and quantum
cryptography \cite{Phoenix-Contemporary-1995,Lo-book-1998,Zeilinger-book-2000,Gisin-RMP-2002,Loepp-book-2006}. For an ensemble $\{\rho_{j},\eta_{j}\}_{j=1}^{n}$ with $\{\rho _{j}\}_{j=1}^{n}$ quantum states and $\{%
\eta_{j}\}_{j=1}^{n}$ a probability distribution, two persons, Alice chooses one $\rho_{j}$ with probability $\eta_{j}$ and sends it to Bob, Bob performs a POVM $E=\{E_{j}\}_{j=1}^{n}$ to determines which $\eta_{j}$ he has received. Bob's maximal success probability then is
\begin{eqnarray}
P^{\text{opt}}(\{\rho_{j},\eta_{j}\}_{j=1}^{n})=\text{max}_{E}\sum_{j}\eta _{j}\text{tr}%
(E_{j}\rho _{j}),  \nonumber
\end{eqnarray}
where max is over all POVMs. When $n=2$, the analytical formula of $P^{\text{opt}}$ and the optimal measurements have been found, but the formulas of $P^{\text{opt}}$ and the optimal measurements are still not found for general $n>2$.

As a suboptional scheme, LSM is a significant method to discriminate quantum states \cite{Belavkin-1975,Belavkin-Stochastics-1975,Holevo-1978,Hausladen-JMO-1994,Wootters-PRA-1996,Wootters-PRL-1991,Eldar-IEEE-2001}.
An LSM has the construction as follows.  let
\begin{eqnarray}
\rho_{\text{out}}=\sum_{j=1}^{n}\eta_{j}\rho_{j},  \label{eq3.1}
\end{eqnarray}
then the LSMs $M=\{M_{j}^{\text{lsm}}\}_{j=1}^{n}$ are given by the measurement operators
\begin{eqnarray}
M_{j}^{\text{lsm}}=\eta _{j}\rho _{\text{out}}^{-\frac{1}{2}}\rho _{j}\rho
_{\text{out}}^{-\frac{1}{2}}.   \label{eq3.2}
\end{eqnarray}
As a result, the minimal error probability of $M$ is given by
\begin{eqnarray}
P^{\text{lsm}}(\{\rho_{j},\eta_{j}\}_{j=1}^{n})=1-\sum_{j}\eta _{j}\text{tr}%
(M_{j}^{\text{lsm}}\rho _{j}).  \label{eq3.3}
\end{eqnarray}
Concerning the relationship between $C_{T,1/2}$ and LSM, we have the following
Theorem 3 and Theorem 4.

\emph{Theorem 3.} Let $E=\{E_{j}\}_{j=1}^{n}$ be a POVM on the Hilbert space $H$
and $\rho $ a state in $H$. We have
\begin{eqnarray}
C_{T,1/2}(\rho ,E)=2P^{\text{lsm}}(\{\rho_{j},\eta_{j}\}_{j=1}^{n}),  \label{eq3.4}
\end{eqnarray}
where
\begin{eqnarray}
\eta _{j}&=&\text{tr}(\rho E_{j}),   \label{eq3.5} \\
\rho _{j}&=&\eta _{j}^{-1}\sqrt{\rho }E_{j}\sqrt{\rho }.  \label{eq3.6}
\end{eqnarray}
\emph{Proof.} According to Eqs. (\ref{eq3.1}), (\ref{eq3.2}) and (\ref{eq3.6}), we have
\begin{eqnarray}
\rho _{\text{out}}&=&\sum_{j=1}^{n}\eta _{j}\rho _{j}=\rho .  \nonumber \\
M_{j}^{\text{lsm}}&=&\eta _{j}\rho _{\text{out}}^{-\frac{1}{2}}\rho _{j}\rho
_{\text{out}}^{-\frac{1}{2}}=E_{j}.    \nonumber \\
\end{eqnarray}
Consequently, we obtain
\begin{eqnarray}
&&P^{\text{lsm}}(\{\rho_{j},\eta_{j}\}_{j=1}^{n})  \nonumber \\
&=&1-\sum_{j}\eta _{j}\text{tr}(M_{j}^{\text{lsm}}\rho _{j})   \nonumber \\
&=&1-\sum_{j}\text{tr}(E_{j}\sqrt{\rho }E_{j}\sqrt{\rho })   \nonumber \\
&=&1-\sum_{j}\text{tr}[(\sqrt{E_{j}}\sqrt{\rho }\sqrt{E_{j}})^{2}]  \nonumber \\
&=&\frac{1}{2}C_{T,1/2}(\rho,E).   \nonumber
\end{eqnarray}
which completes the proof.
$\hfill\blacksquare$

Theorem 3 shows that the minimal error probability of the least square
measurement state discrimination is given by the POVM coherence $C_{T,1/2}(\rho,E)$ based on Tsallis
relative entropy. If $\rho$ is POVM incoherent with respect to POVM $E=\{E_{j}\}_{j=1}^{n},$
i.e., $C_{T,1/2}(\rho ,E)=0,$ then $P^{\text{lsm}}(\{\rho_{j},\eta
_{j}\}_{j=1}^{n})=0,$ which means that $\{\rho_{j},\eta
_{j}\}_{j=1}^{n}$ can be perfectly discriminated by the least square
measurement $E=\{E_{j}\}_{j=1}^{n}.$

From Theorem 3, we also see that for a given POVM $E=\{E_{j}\}_{j=1}^{n}$ and a
state $\rho$, there exists an ensemble $\{\rho_{j},\eta_{j}\}_{j=1}^{n}
$ such that Eq. (\ref{eq3.4}) holds. Conversely, we
may ask for a given ensemble $\{\rho_{j},\eta _{j}\}_{j=1}^{n}$,
whether a POVM $E=\{E_{j}\}_{j=1}^{n}$ and a state $\rho $ exist so
that Eq. (\ref{eq3.4}) holds. Theorem 4 below shows
that this is true, which can be verified similar to the proof of Theorem 3.

\emph{Theorem 4.} For a given ensemble $\{\rho_{j},\eta_{j}\}_{j=1}^{n}$ on
the Hilbert space $H,$ there exist a quantum state $\rho $ and a POVM $%
E=\{E_{j}\}_{j=1}^{n}$ such that
\begin{eqnarray}
2P^{\text{lsm}}(\{\rho_{j},\eta_{j}\}_{j=1}^{n})=C_{T,1/2}(\rho ,E), \label{eq3.7}
\end{eqnarray}
where
\begin{eqnarray}
\rho &=&\sum_{j=1}^{n}\eta_{j}\rho _{j}, \label{eq3.8}  \\
E_{j}&=&\eta_{j}\rho ^{-\frac{1}{2}}\rho _{j}\rho ^{-\frac{1}{2}}.  \label{eq3.9}
\end{eqnarray}

We remark that when the POVM $E=\{E_{j}\}_{j=1}^{n}$ is a rank-one projective
measurement $E=\{|j\rangle \langle j|\}_{j=1}^{d}$, both the Theorem 3 and
Theorem 4 reduce to the corresponding conclusions in the theory of conventional coherence \cite{Xiong-2018-PRA}.

\section{Uncertainty relations given by $C_{r}$}
The entropic uncertainty relations play a central role in quantum cryptographic protocols \cite{Coles-RMP-2017}.
In this section, we establish an entropic uncertainty relation given by the relative entropy of the POVM coherence $%
C_{r}(\rho,E)$.

\emph{Theorem 5.} Let $E=\{E_{j}\}_{j=1}^{n}$ and $F=\{F_{k}\}_{k=1}^{m}$ be two POVMs
on the Hilbert space $H$ and $\rho$ a state in $H$. We have
\begin{eqnarray}
C_{r}(\rho ,E)+C_{r}(\rho ,F)\geq 2[\log _{2}\frac{1}{c}-S(\rho )],  \label{eq4.1}
\end{eqnarray}
where $c=\max_{jk}c_{jk}$, $c_{jk}=||\sqrt{E_{j}}\sqrt{F_{k}}||$, $||\cdot ||$ denotes the operator norm (the maximal singular value).

\emph{Proof.} Rewrite Eq. ({\ref{eq1.3}) as
\begin{eqnarray}
C_{r}(\rho ,E)=S(\{\text{tr}(E_{j}\rho )\}_{j=1}^{n}) \ \ \ \ \ \ \ \ \ \  \ \ \ \ \ \ \ \ \ \  \ \ \ \ \    \nonumber \\
+\sum_{j=1}^{n}\text{tr}(E_{j}\rho )S(\sqrt{E_{j}%
}\rho \sqrt{E_{j}}/\text{tr}(E_{j}\rho ))-S(\rho ).  \label{eq4.3}
\end{eqnarray}
By using the result \cite{Krishna-IJS-2002} that
\begin{eqnarray}
S(\{\text{tr}(E_{j}\rho )\}_{j=1}^{n})+S(\{\text{tr}(F_{k}\rho )\}_{k=1}^{m})\geq
2\log _{2}\frac{1}{c}, \label{eq4.4}
\end{eqnarray}
we then can prove Theorem 5. $\hfill\blacksquare$

Theorem 5 can be improved when $c$ is improved by $c'$ as
in \cite{Tomamichel-PhDthesis-2012, Coles-PRA-2014}.
\begin{eqnarray}
c'=\text{min}\{\text{max}_{k}||\sum_{j}E_{j}F_{k}E_{j}||,\text{max}_{j}||\sum_{k}F_{k}E_{j}F_{k}||\}, \nonumber
\end{eqnarray}
Replacing $c$ by $c'$ in Eq. (\ref{eq4.1}), it yields a stronger uncertainty relation \cite{Tomamichel-PhDthesis-2012, Coles-PRA-2014}.

There is an operational interpretation of $C_{r}(\rho ,E)$ \cite{Brub-2021-PRA}: $C_{r}(\rho ,E)$ quantifies
the private randomness generated by the POVM $E$ on the state with respect to an
eavesdropper holding optimal side information about the measured state. In this case Eq. (\ref{eq4.1}) sets a
tradeoff constraint between the POVMs $E$ and $F.$  When POVMs $E$ and $F$ are rank-one
projective measurements, Eq. (\ref{eq4.1}) reduces the case of conventional coherence \cite{Ma-EPL-2019}.

For any normalized pure state $|\psi \rangle ,$  Theorem 5 gives rise to
\begin{eqnarray}
C_{r}(\psi,E)+C_{r}(\psi,F)\geq \log _{2}\frac{1}{c},
\end{eqnarray}
where the lower bound is no longer state-dependent.

\section{Averages of $C_{r}$, $C_{T,\alpha }$ and $C_{l_{1}}$ over random pure quantum states}

Random pure quantum states offer new insights into various phenomena in quantum physics and quantum information theory \cite{Collins-JMP-2016}.
For the space of $d$-dimensional pure states, there exists a unique measure $d\psi$
induced from the uniform Haar measure $d\mu(U)$ on the unitary group $U(d)$ \cite{Collins-JMP-2016}. This amounts to saying that any random pure state can be seen as a unitary matrix $U\in U(d)$ performing on a fixed pure state $\psi_{0}$. In the following we always adopt this measure for random pure states \cite{Singh-PRA-2016,Zhang-JPA-2017,Wu-JPA-2020}. The average value of the function $g(\psi)$ over all pure states $\psi$ then is defined as
\begin{eqnarray}
\int g(\psi)d\psi=\int_{U(d)}g(U\psi_{0})d\mu(U)   \nonumber
\end{eqnarray}

The average of conventional coherence for pure quantum states has been extensively studied
\cite{Singh-PRA-2016,Zhang-JPA-2017,Wu-JPA-2020}. In \cite{Luo-PLA-2019} the authors presented the conventional coherence average over all orthonormal bases (rank-one projective measurements). In this section, we investigate the averages of $C_{r}(\rho ,E)$, $C_{T,\alpha
}(\rho ,E)$ and $C_{l_{1}}(\rho ,E)$ with respect to POVM $E$ for random pure states. We have Theorem 6 below.

\emph{Theorem 6.} The averages of $C_{r}(\psi ,E)$, $C_{T,\alpha }(\psi ,E)$
and $C_{l_{1}}(\psi ,E)$ for random pure states $\psi$ have the properties
\begin{eqnarray}
\int d\psi C_{r}(\psi ,E)=-\frac{1}{d\ln 2}  \ \ \ \ \ \ \ \ \ \ \ \ \ \ \ \ \ \ \ \ \ \ \ \ \ \ \ \ \ \ \  \nonumber \\
\cdot\sum_{j=1}^{n}\sum_{k=1}^{d}(\Pi
_{l\neq k}\frac{1}{\lambda _{j,k}-\lambda _{j,l}})\lambda _{j,k}^{d}(\ln
\lambda _{j,k}-\sum_{m=2}^{d}\frac{1}{m}), \label{eq5.4}  \\
\int d\psi C_{T,\alpha }(\psi ,E)=\frac{1}{\alpha -1}\{%
\sum_{j=1}^{n}B(E_{j},\frac{1}{\alpha })-1\}, \label{eq5.5}  \ \ \ \ \  \\
\int d\psi C_{l_{1}}(\psi ,E)\leq \sum_{j\neq k}(\frac{B(E_{j},\frac{p_{jk}%
}{2})}{p_{jk}}+\frac{B(E_{k},\frac{q_{jk}}{2})}{q_{jk}}),   \ \ \label{eq5.6}
\end{eqnarray}
where $\{\lambda _{j,l}\}_{l=1}^{d}$ are the eigenvalues of $E_{j},$
\begin{eqnarray}
B(E_{j},\beta )=\frac{\Gamma (d)\Gamma (1+\beta )}{\Gamma (d+\beta )}%
\sum_{k=1}^{d}(\Pi _{l\neq k}\frac{1}{\lambda _{j,k}-\lambda _{j,l}})\lambda
_{j,k}^{d+\beta -1}, \nonumber \\
\label{eq5.7}
\end{eqnarray}
$\beta >0,$ $\Gamma (\cdot )$ is the Gamma function, and $%
p_{jk}>1$, $q_{jk}>1$, $1/p_{jk}+1/q_{jk}=1$ for any $j\neq k.$ $B(E_{j},\beta )$ is discussed in Appendix A of \cite{Zhang-QIP-2018}.

\emph{Proof.} Let $A$ be an operator on $H$ with eigendecomposition,
\begin{eqnarray}
A=\sum_{j=1}^{d}\lambda _{j}|\varphi _{j}\rangle \langle \varphi _{j}|,  \label{eq5.8}
\end{eqnarray}
where $\{\lambda _{j}\}_{j=1}^{d}$ are the real eigenvalues and $\{|\varphi _{j}\rangle
\}_{j=1}^{d}$ constitute an orthonormal basis of $H$. Then one has \cite{Jones-JPA-1991}
\begin{eqnarray}
&&\int f(\langle \psi |A|\psi \rangle )d\psi \nonumber \\
&=&\Gamma (d)\sum_{j=1}^{d}(\Pi
_{k\neq j}\frac{1}{\lambda _{j}-\lambda _{k}})\mathcal{R}_{d-1}[f](\lambda
_{j})
\label{eq5.9}
\end{eqnarray}
for any function $f(\cdot )$, where $\mathcal{R}_{d-1}[f]$ is the
Riemann-Liouville fractional integration defined by
\begin{eqnarray}
\mathcal{R}_{\mu }[f](u)=\frac{1}{\Gamma (\mu )}\int_{0}^{u}f(w)(u-w)^{\mu
-1}dw,\text{Re} \mu >0.  \label{eq5.10}
\end{eqnarray}
Note that \cite{Erdelyi-1954-book}
\begin{eqnarray}
&&\mathcal{R}_{\mu }[w^{\nu -1}\ln w](u)  \nonumber \\
&=&\frac{\Gamma (\nu )}{\Gamma (\nu
+\mu )}u^{\mu +\nu -1}(\ln u+\Psi (\nu )-\Psi (\mu +\nu )),  \label{eq5.11}
\end{eqnarray}
with Re$\nu >0,$ Re$\mu >0$ and $\Psi (\nu )=\frac{d}{d\nu }\ln \Gamma (\nu )$
the digamma function. Set $\mu =d-1$ and $\nu =2$. Using the properties $%
\Gamma (d+1)=d!$ and $\Psi (\mu +1)=\Psi (\mu )+1/\mu ,$ we get
\begin{eqnarray}
\mathcal{R}_{d-1}[w\ln w](u)=\frac{1}{d!}u^{d}(\ln u-\sum_{m=2}^{d}\frac{1}{%
m}). \label{eq5.12}
\end{eqnarray}
Since
$$
\int d\psi C_{r}(\psi ,E)=-\sum_{j=1}^{n}\int d\psi \langle \psi
|E_{j}|\psi \rangle \log _{2}\langle \psi |E_{j}|\psi \rangle,
$$
we get (\ref{eq5.4}) from (\ref{eq5.9}) and (\ref{eq5.12}).

Note that \cite{Erdelyi-1954-book}
\begin{eqnarray}
\mathcal{R}_{\mu }[w^{\nu -1}](u)=\frac{\Gamma (\nu )}{\Gamma (\nu +\mu )}%
u^{\mu +\nu -1}, \label{eq5.13}
\end{eqnarray}
for Re$\nu >0$ and Re$\mu >0.$ Set $\mu =d-1$ and $\nu -1=\beta$. We have
\begin{eqnarray}
\mathcal{R}_{d-1}[w^{\beta}](u)=\frac{\Gamma (1+\beta)}{\Gamma (d+\beta)}u^{d+\beta-1}.  \label{eq5.14}
\end{eqnarray}
By definition,
$$
\int d\psi C_{T,\alpha }(\psi ,E)=\frac{1}{\alpha -1}[\sum_{j=1}^{n}\int
d\psi \langle \psi |E_{j}|\psi \rangle ^{\frac{1}{\alpha }}-1].
$$
Hence we obtain (\ref{eq5.5}) from (\ref{eq5.9}) and (\ref{eq5.14}).

Moreover, from
\begin{eqnarray}
C_{l_{1}}(\psi ,E)&=&\sum_{j\neq k} \sqrt{\langle \psi
|E_{j}|\psi \rangle \langle \psi |E_{k}|\psi \rangle}, \nonumber \\
\int d\psi C_{l_{1}}(\psi ,E)&=&\sum_{j\neq k}\int d\psi \sqrt{\langle \psi
|E_{j}|\psi \rangle \langle \psi |E_{k}|\psi \rangle},  \nonumber
\end{eqnarray}
we have
\begin{eqnarray}
C_{l_{1}}(\psi ,E)\leq \sum_{j\neq k}(\frac{\langle \psi |E_{j}|\psi
\rangle ^{\frac{p_{jk}}{2}}}{p_{jk}}+\frac{\langle \psi |E_{k}|\psi \rangle
^{\frac{q_{jk}}{2}}}{q_{jk}}), \label{eq5.15}
\end{eqnarray}
where we have used the Young inequality,
\begin{eqnarray}
ab\leq \frac{a^{p_{jk}}}{p_{jk}}+\frac{b^{q_{jk}}}{q_{jk}}  \label{eq5.16}
\end{eqnarray}
for $a\geq 0$, $b\geq 0,$ $p_{jk}>1$, $q_{jk}>1$ and $\frac{1}{p_{jk}}+\frac{1}{q_{jk}}%
=1.$
Taking into account (\ref{eq5.15}), (\ref{eq5.9}) and (\ref{eq5.14}), one gets (\ref{eq5.6}). $\hfill\blacksquare$

Note that if the eigenvalues $\{\lambda _{j,l}\}_{l=1}^{d}$ in (\ref{eq5.4}) and (\ref{eq5.7}) are degenerate,
the problem can be still handled via the standard trick of taking the limits when the eigenvalues
approach pairwise \cite{Jones-JPA-1991}, see also \cite{Zhang-QIP-2018} for (\ref{eq5.7}).
In particular, if $p_{jk}=q_{jk}=2$ for any $j\neq k$ in (\ref{eq5.15}), we get
\begin{eqnarray}
C_{l_{1}}(\psi ,E)\leq n-1.  \label{eq5.17}
\end{eqnarray}

We remark that when the POVM $E$ is a rank-one projective
measurement $E=\{|j\rangle \langle j|\}_{j=1}^{d}$, (\ref{eq5.4}) gives rise to the corresponding result for conventional
coherence \cite{Singh-PRA-2016},
\begin{eqnarray}
\int d\psi C_{r}(\psi ,\{|j\rangle \langle j|\}_{j=1}^{d})=\frac{1%
}{\ln 2}\sum_{m=2}^{d}\frac{1}{m}.\nonumber
\end{eqnarray}

To compare (\ref{eq5.5}) with the one of conventional
coherence, we consider the case of $\alpha =1/2$.
We see that $\sum_{k=1}^{d}(\Pi _{l\neq k}\frac{1}{\lambda _{j,k}-\lambda _{j,l}})\lambda
_{j,k}^{d+1}$ is a homogeneous symmetric polynomial in $\{\lambda
_{j,k}\}_{k=1}^{d}$ of degree 2, which
can be expressed in terms of elementary symmetric polynomials. Due to the homogenity,
there exist constants $C_{1}$ and $C_{2}$ such that
\begin{eqnarray}
&&\sum_{k=1}^{d}(\Pi _{l\neq k}\frac{1}{\lambda _{j,k}-\lambda _{j,l}}%
)\lambda _{j,k}^{d+1} \nonumber \\
&=&C_{1}(\sum_{k=0}^{d}\lambda
_{j,k})^{2}+C_{2}\sum_{k<l}\lambda _{j,k}\lambda _{j,l}.
\end{eqnarray}
Since
\begin{eqnarray}
(\sum_{k=0}^{d}\lambda _{j,k})^{2}=\sum_{k=0}^{d}\lambda
_{j,k}^{2}+2\sum_{k<l}\lambda _{j,k}\lambda _{j,l},
\end{eqnarray}
there further exist constants $C_{3}$ and $C_{4}$ such that
\begin{eqnarray}
&&\sum_{k=1}^{d}(\Pi _{l\neq k}\frac{1}{\lambda _{j,k}-\lambda _{j,l}}%
)\lambda _{j,k}^{d+1}  \nonumber \\
&=&C_{3}(\sum_{k=0}^{d}\lambda
_{j,k})^{2}+C_{4}\sum_{k=1}^{d}\lambda _{j,k}^{2}.
\end{eqnarray}
Set $\lambda _{j,1}=1$ and $\{\lambda _{j,k}=0\}_{k=2}^{d}$, we get $C_{3}+C_{4}=1$.
Let $\lambda _{j,1}=1$, $\lambda _{j,2}=1/2$ and $\{\lambda _{j,k}=0\}_{l=k}^{d}$, we have $\frac{9}{4}C_{3}+\frac{5}{4}C_{4}=\frac{7}{4}.$ Consequently, we obtain $C_{3}=C_{4}=\frac{1}{2}$ and
\begin{eqnarray}
&&\int d\psi C_{T,\frac{1}{2}}(\psi ,E)  \nonumber \\
&=&2\{1-\frac{1}{d(d+1)}\sum_{j=1}^{n}[(\text{tr}E_{j})^{2}+\text{tr}(E_{j}{}^{2})]\}.  \label{eq5.19}
\end{eqnarray}

Hence, when the POVM $E$ is a rank-one projective
measurement $E=\{|j\rangle \langle j|\}_{j=1}^{d}$, (\ref{eq5.19}) recovers the result in \cite{Wu-JPA-2020},
$\int d\psi C_{T,\frac{1}{2}}(\psi ,\{|j\rangle \langle j|\}_{j=1}^{d})=2({d-1})/({d+1})$.

\section{Summary}
The conventional quantum coherence is defined with respect to an orthonormal basis,
while the generalized quantum coherence studied is defined with respect to general POVM settings. We
have investigated the properties of three well-defined POVM coherence measures, the $C_{r}(\rho ,E)$,
$C_{T,\alpha }(\rho ,E)$ and $C_{l_{1}}(\rho ,E)$. We have provided the upper bounds of $C_{l_{1}}(\rho,E)$,
the operational interpretation for $C_{T,1/2}(\rho,E)$, the uncertainty relations given by $C_{r}(\rho ,E)$,
and calculated the averages of $C_{r}(\rho ,E)$, $C_{T,\alpha }(\rho ,E)$ and $C_{l_{1}}(\rho,E)$ over random pure states.
These results will strengthen the necessity of the concept of POVM coherence, and highlight the potential applications of these POVM coherence measures.

\section*{ACKNOWLEDGMENTS}
This work was supported by the Chinese Universities Scientific Fund under Grant No. 2452021067, the National Science Foundation of China under Grant Nos. 11675113, 11971140 and 12075159, the Key Project of Beijing Municipal Commission of Education (KZ201810028042), Beijing Natural Science Foundation (Z190005), Academy for Multidisciplinary Studies, Capital Normal University; Shenzhen Institute for Quantum Science and Engineering, Southern University of Science and Technology (No. SIQSE202001) and the Academician Innovation Platform of Hainan Province.

%

%\bibliographystyle{apsrev4-1}
%\bibliography{POVMcoherence}

\end{document}